\documentclass[11pt]{article}
\usepackage{latexsym}  
\usepackage{amssymb}
\usepackage{graphicx}
\usepackage{amsmath}

\begin{document}

\begin{center}

{\Large\bf Neutrino Mass Matrix Model \\
with Only Three Adjustable Parameters}

\vspace{4mm}

{\bf Yoshio Koide$^a$ and Hiroyuki Nishiura$^b$}

${}^a$ {\it Department of Physics, Osaka University, 
Toyonaka, Osaka 560-0043, Japan} \\
{\it E-mail address: koide@kuno-g.phys.sci.osaka-u.ac.jp}

${}^b$ {\it Faculty of Information Science and Technology, 
Osaka Institute of Technology, 
Hirakata, Osaka 573-0196, Japan}\\
{\it E-mail address: nishiura-hiroyuki-hp@alumni.osaka-u.ac.jp}
\date{\today}
\end{center}

\vspace{3mm}

\begin{abstract}
Stimulated by a successful quark mass matrix model based on U(3)$\times$U(3)$'$ 
family symmetry, a phenomenological neutrino mass matrix 
for the Majorana neutrinos $(\nu_L, \nu_R^c, N_L, N_R^c)$ is proposed. 
The model has only three adjustable parameters. 
Nevertheless, the model gives reasonable predictions 
for the  neutrino masses, mixings, and CP violating phases in the neutrino mixing matrix.
\end{abstract}

PCAC numbers:  
  11.30.Hv, 
  12.60.-i, 
  14.60.Pq,  


\vspace{3mm}

{\large\bf 1. \ Introduction }
 
 The greatest concern in the flavor physics is how to understand
the origin of the observed structures of  masses 
and mixings of quarks and leptons. 
Recently, we have proposed a quark mass matrix model 
based on the U(3)$\times$U(3)$'$ flavor symmetry 
\cite{K-N_PRD15R, K-N_MPLA16}, which originates from the following 
$2\times 2$ blocks mass matrix model:
$$
(\bar{f}_L^i \ ,\ \bar{F}_L^\alpha ) 
\left(
\begin{array}{cc}
(0)_i^{\ j}  &  (\Phi_f)_i^{\ \beta}  \\
(\bar{\Phi}_{f})_\alpha^{\ j} & -(S_f)_\alpha^{\ \beta} 
\end{array} \right) 
 \left(
\begin{array}{c}
f_{Rj} \\
F_{R\beta}
\end{array} 
\right) .
\eqno(1.1)
$$
Here,
we consider hypothetical heavy fermions $F_\alpha$ 
($\alpha=1,2,3$), which belong to $({\bf 1}, {\bf 3})$ of
 U(3)$\times$U(3)$'$, 
in addition to quarks and charged leptons $f_i=(u_i, d_i, e_i)$ ($i=1,2,3$) 
which belong to  $({\bf 3}, {\bf 1})$.
The fields $\Phi_f$ and $S_f$ are scalars which belong to 
$({\bf 3}, {\bf 3}^*)$ and $( {\bf 1}, {\bf 8}+{\bf 1})$ 
of U(3)$\times$U(3)$'$, respectively.
According to a seesaw-like mechanism,
we obtain quarks and charged lepton mass matrices
$$
(M_f)_i^{\ j} = \langle (\Phi_f)_i^{\ \alpha} \rangle 
\langle (S_f^{-1})_\alpha^{\ \beta} \rangle 
 \langle (\Phi_f)_\beta^{\ j} \rangle , 
 \eqno(1.2)
$$
where $\alpha$ and $\beta$ are indexes of U(3)$'$ and 
$i$ and $j$ are indexes of U(3).
(Here, exactly speaking, the matrix $(M_f)_i^{\ j}$ in Eq.(1.2) 
represents the Yukawa coupling constant 
 $(Y_f)_i^{\ j}$ of the fermion $f$. 
However,  for convenience, we will call it as "mass matrix". )

Furthermore, we assume the vacuum expectation values (VEVs) of 
those scalars as follows:
$$
\langle {S}_f^{-1} \rangle = [v_{S} ({\bf 1} + b_f  X_3)]^{-1} 
= v_{S}^{-1} ({\bf 1} + a_f  X_3), 
\eqno(1.3)
$$
where ${\bf 1}$ and $X_3$ are defined by
$$
{\bf 1} = \left( 
\begin{array}{ccc}
1 & 0 & 0 \\
0 & 1 & 0 \\
0 & 0 & 1 
\end{array} \right) , \ \ \ \ \ 
X_3 = \frac{1}{3} \left( 
\begin{array}{ccc}
1 & 1 & 1 \\
1 & 1 & 1 \\
1 & 1 & 1 
\end{array} \right) ,  
\eqno(1.4)
$$
and $a_f$ and $b_f$ are complex parameters with the relation
$$
a_f = - \frac{b_f}{1+ b_f} .
\eqno(1.5)
$$
On the other hand, the VEV form of $\langle \Phi_f \rangle$
are chosen as
$$
\langle \Phi_f \rangle = v_\Phi\, {\rm diag}( z_1 e^{i\phi^f_1},   
 z_2 e^{i\phi^f_2}, z_3 e^{i\phi^f_3}) ,
\eqno(1.6)
$$
where
$$
z_i = \frac{ \sqrt{m_{ei}} }{\ \sqrt{ 
m_e + m_\mu + m_\tau} } .
\eqno(1.7)
$$
Here, 
$m_{ei}=(m_e, m_\mu. m_\tau)$ are charged lepton masses.                       

The mass matrix form (1.2) with the form (1.3) is known 
as the ``democratic seesaw mass matrix" form \cite{democratic}.
The term ``democratic" was named by Jarlskog \cite{Jarlskog}.
In this model, parameters which we can adjust are only 
the generation-independent parameters $a_f$ ($f= u$ and $d$ ).

Under this mass matrix model (1.2), we have successfully 
obtained \cite{K-N_PRD15R}  a unified description of 
quark masses and mixing by using the observed charged 
lepton masses.  
The models for quarks  were proposed in the previous century, but 
we consider that the basic idea should be still inherited to  
the neutrino mass matrix model.

A naive extension of the quark mass matrix model (1.1) 
to the $2 \times 2$ blocks mass matrix for neutrino $\nu$ 
and heavy neutrino  $N$ will be as follows:
$$
(\bar{\nu}_L^i \ ,\ \bar{N}_L^\alpha ) 
\left(
\begin{array}{cc}
(0)_{ij}  &  (\Phi_\nu)_{i\beta}  \\
(\bar{\Phi}_\nu)_{\alpha j}) & -(S_\nu)_{\alpha \beta}
\end{array} \right) 
 \left(
\begin{array}{c}
(\nu_L^c)^j \\
(N_L^c)^\beta
\end{array} \right) , 
\eqno(1.8)
$$
which leads to
$$
(M_\nu)_{ij} = \langle (\Phi_\nu)_i^{\ \alpha} \rangle 
\langle (S_\nu^{-1})_{\alpha\beta} \rangle 
 \langle (\Phi_\nu)^\beta_{\ j} \rangle . 
 \eqno(1.9)
$$
However, this neutrino mass matrix (1.9) is too simplified. 
We think that the Majorana mass terms must be taken into consideration.
 
In this paper, we consider a $4\times 4$ blocks mass matrix 
\cite{YK_PRD96}
for the neutrino states $(\nu_L, \nu_R^c, N_L, N_R^c)$ 
($f^c$ denote a charge conjugate state of a fermion $f$).
The explicit form will be given in the next section. 
Our model is a three parameter model, and those parameters
are fixed by three of the four  observed  values in the  neutrino oscillation data.
Thereby, we can make good fitting for the rest observed values, 
and we predict the $CP$ violation phase factor which
will be  soon observed rigidly. 

\vspace{5mm}. 

{\large\bf  2. \ Majorana neutrino mass matrix }
      
In the present paper, we assume the following mass 
matrix: 
$$
((\bar{\nu}_L)^\circ \ ,\ (\bar{\nu}_R^c)_\circ \ ,\ 
(\bar{N}_L)^\bullet \ ,\ (\bar{N}_R^c)_\bullet )
$$
$$
\times 
\left(
\begin{array}{cccc} 
 ( 0 )_{\circ\circ}  &  ( {\bf 1} )_\circ^{\ \circ}  & 
 ( \Phi )_\circ^{\ \bullet}  &  ( 0 )_{\circ \bullet} \\[0.02in]  
( {\bf 1} )^{\circ}_{\ \circ}  &  ( 0 )^{\circ\circ}  &  
( {\bf 1} )^{\circ\bullet}  &   ( \Phi)^{\circ}_{\ \bullet} \\[0.05in] 
\Phi^\bullet_{\ \circ}  &  ( {\bf 1} )^{\bullet\circ}   &
    S^{\bullet\bullet}  &  S^\bullet_{\ \bullet} \\[0.02in]  
( 0 )_{\bullet\circ}  &  ( 0 )_\bullet^{\ \circ}  & 
 S_\bullet^{\ \bullet}  & ({\bf 1} )_{\bullet\bullet} \\[0.02in] 
 \end{array} \right) \ \ \ \ 
 \left(
 \begin{array}{c}
 (\nu_L^c)^\circ  \\[0.02in] 
 (\nu_R)_\circ   \\[0.05in] 
 %
  %
  (N_L^c)^\bullet   \\[0.02in] 
  (N_R)_\bullet   \\[0.02in] 
  \end{array}
  \right) . 
\eqno(2.1)
$$
Here, for convenience, we put $\circ$ for  indexes $i, j, \cdots$ 
and $\bullet$ for $\alpha, \beta, \cdots$. 
And also, for convenience, we dropped the symbols 
``$\langle$" and ''$\rangle$".
In the mass matrix, the element $( 0 )$ shows an empty element.
$\Phi_\circ^{\ \bullet}$, $\Phi^\circ_{\ \bullet}$, 
$S_\bullet^{\ \bullet}$ and $S^\bullet_{\ \bullet}$ were  
already introduced in the quark mass matrix model 
\cite{K-N_PRD15R, K-N_MPLA16}. 

As characteristic Majorana mass terms, we have 
introduced $({\bf 1})^{\circ\bullet}$, $({\bf 1})^{\bullet\circ}$,
$S^{\bullet\bullet}$ and $({\bf 1})_{\bullet\bullet}$, so that we take the same form as in Eq.(1.9).  
Here,  the term $S^{\bullet\bullet}$ is a Majorana version of the Dirac 
mass terms  $S_\bullet^{\ \bullet}$ and $S^\bullet_{\ \bullet}$. 

Since we demand that the number of free parameters in the model 
is as small as possible, 
for the other Majorana mass terms $({\bf 1})^{\circ\bullet}$, 
 $({\bf 1})^{\bullet\circ}$ and $({\bf 1})_{\bullet\bullet}$, 
we assumed that those are structureless, e.g. a unit matrix. 
Also, we assumed unit matrix forms for mass terms with lower 
energy scale  
$({\bf 1})_\circ^{\ \circ}$ and $({\bf 1})^\circ_{\ \circ}$.
We will refer the mass matrix  given in Eq.(2.1)
as $M_{4\times 4}$.

In the assignments of the scalars in the mass matrix 
$M_{4\times 4}$, there is no theoretical inevitability. 
We demand that the main term takes the familiar form (1.8),  
and that the second term takes a simple and plausible form. 

We obtain the neutrino mass matrix $M_\nu$ from the 
generalized mass matrix $M_{4\times 4}$ by using  the following seesaw approximation as follows:
$$
M_{4\times 4}\ \Rightarrow \ M_{3\times 3} = \left( 
\begin{array}{ccc}
( 0 )_{\circ\circ} & {\bf 1}_\circ^{\ \circ} & \Phi_\circ^{\ \bullet}\\
{\bf 1}^\circ_{\ \circ} & ( 0 )^{\circ\circ} & {\bf 1}^{\circ\bullet} \\
\Phi^\bullet_{\ \circ} & {\bf 1}^{\bullet\circ} & S^{\bullet\bullet}
\end{array} \right)
-\left( \begin{array}{c}
( 0 )_{\circ\bullet} \\
( 0 )^\circ_{\ \bullet} \\
S^\bullet_{\ \bullet} 
\end{array} \right) 
 ( {\bf 1}_{\bullet\bullet} )^{-1} 
\left( ( 0 )_{\bullet\circ} \ ,\ 
( 0 )_\bullet^{\ \circ} \ ,\ 
 S_\bullet^{\ \bullet} \right)
 $$
 $$
 = \left( \begin{array}{ccc}
 ( 0 )_{\circ\circ} & {\bf 1}_\circ^{\ \circ} & \Phi_\circ^{\ \bullet} \\
 {\bf 1}^\circ_{\ \circ} & ( 0 )^{\circ\circ} & {\bf 1}^{\circ\bullet} \\
 \Phi^\bullet_{\ \circ} & {\bf 1}^{\bullet\circ} &  S^{\bullet\bullet} -
S^\bullet_{\ \bullet} ({\bf 1}_{\bullet\bullet})^{-1}S_\bullet^{\ \bullet}
 \end{array} \right) ,
 \eqno(2.2)
 $$
$$
M_{3\times 3}\ \Rightarrow \ M_{2\times 2} = \left( 
\begin{array}{cc}
-\Phi_\circ^{\ \bullet} ((S_{eff})^{-1})_{\bullet\bullet}
 \Phi^\bullet_{\ \circ} & 
{\bf 1}_\circ^{\ \circ} -\Phi_\circ^{\ \bullet} 
(S_{eff})^{-1})_{\bullet\bullet}{\bf 1}^{\bullet\circ} \\
{\bf 1}^\circ_{\ \circ} -{\bf 1}^{\circ\bullet}
((S_{eff})^{-1})_{\bullet\bullet} \Phi^\bullet_{\ \circ} &
-{\bf 1}^{\circ\bullet} ((S_{eff})^{-1})_{\bullet\bullet}
 {\bf 1}^{\bullet\circ}
\end{array} \right) ,
\eqno(2.3)
$$
where
$$
(S_{eff})^{\bullet\bullet} \equiv  S^{\bullet\bullet} - S^\bullet_{\ \bullet} 
({\bf 1}_{\bullet\bullet} )^{-1} S_\bullet^{\ \bullet}   .
\eqno(2.4)
$$
Then, we obtain the neutrino mass matrix
$$
M_{2\times 2} \ \Rightarrow \ (M_\nu)_{\circ\circ} = 
-\Phi_\circ^{\ \bullet} (S_{eff}^{-1})_{\bullet\bullet} 
\Phi^\bullet_{\ \circ} 
$$
$$
+ [{\bf 1}_\circ^{\ \circ} -\Phi_\circ^{\ \bullet} 
((S_{eff})^{-1})_{\bullet\bullet}{\bf 1}^{\bullet\circ}]
({\bf 1}^{\circ\bullet})^{-1} (S_{eff})^{\bullet\bullet}
({\bf 1}^{\bullet\circ})^{-1} [{\bf 1}^\circ_{\ \circ} 
- {\bf 1}^{\circ\bullet} ((S_{eff})^{-1})_{\bullet\bullet}
\Phi^\bullet_{\ \circ}  ].
\eqno(2.5)
$$

This neutrino mass matrix (2.5) is too complicated for a numerical analysis.
Therefore, we take the following approximation:
$$
|{\bf 1}^\circ_{\ \circ} | \gg \left|{\bf 1}^{\circ\bullet} 
((S_{eff})^{-1})_{\bullet\bullet} 
\Phi^\bullet_{\ \circ} \right| ,
\eqno(2.6)
$$
in Eq.(2.5).  
Then, we obtain a simple form
$$
(M_\nu)_{\circ\circ} = 
-\Phi_\circ^{\ \bullet} ((S_{eff})^{-1})_{\bullet\bullet}
 \Phi^\bullet_{\ \circ}+ {\bf 1}_\circ^{\ \circ} 
({\bf 1}^{\circ\bullet})^{-1} (S_{eff})^{\bullet\bullet} 
({\bf 1}^{\bullet\circ})^{-1}
({\bf 1}^\circ_{\ \circ}  ).
\eqno(2.7)
$$

\vspace{5mm}

{\large\bf  3.  \ Numerical estimates}\ \ \ \  

According to Eq.(2.7), we estimate the following mass matrix
$$
M_\nu = k_\nu \{\Phi_\nu ( {\bf 1} + a_{eff} X_3 ) \Phi_\nu + 
\xi ({\bf 1} + a_{eff} X_3 )^{-1} \} . 
\eqno(3.1)
$$
Here, the parameter $a_{eff}$ is defined by 
$\langle {S}_{eff}^{-1} \rangle = v_{S\, eff}^{-1} ({\bf 1} + a_{eff}  X_3)$, 
and $k_\nu$ is an overall factor determined by the VEV scales.
 Since we are interested only in the mass ratios and mixing matrix, 
hereafter, we put $k_\nu = 1$, and    
we use dimensionless expressions for $\Phi_\nu$ given by
$$
\Phi_\nu = {\rm diag}( z_1 e^{i\phi_1}, 
z_2 e^{i\phi_2}, z_3 e^{i\phi_3}) .
\eqno(3.2)
$$
where $\phi_1=\phi_2=\phi_3=0$. 

We have only two (complex) parameters, $a_{eff}$ and $\xi$ in (3.1).
As we discuss later, since we have  only four  neutrino data,
the number of parameters must be smaller than three to avoid 
falling in a mere parameter-fitting model. 
Therefore, one of parameters $a_{eff}$ and $\xi$ must be real. 
In our previous study for the quark mass matrix, we have taken 
the parameter $a_f$ as complex. 
In the present neutrino mass matrix model, too, we assume $a_{eff}$ 
is complex, so that we take $\xi$ as real.  
So we have only three real parameters $a_{eff}=|a_{eff}| e^{i \alpha}$ and 
$\xi$.

Since the neutrino mass matrix $M_\nu$ is a symmetric, i.e. 
$M_\nu^T = M_\nu$, the mass matrix  is diagonalized by 
an unitary matrix $U$ as 
$$
U^T M_\nu  U = D_\nu  \equiv {\rm diag}( m_1, m_2, m_3) ,
\eqno(3.3)
$$
where $m_i$ are the Majorana neutrino masses.

For convenience, hereafter, we use 
the following standard form $U_\nu$ given by
$$
U_\nu 
= \left(
\begin{array}{ccc}
c_{13}c_{12} & c_{13}s_{12} & 
s_{13}e^{-i \delta_{CP}} \\
-c_{23}s_{12}-s_{23}c_{23}s_{13} e^{i\delta_{CP}}
&(c_{23}c_{12}-s_{23}s_{12}s_{13} e^{i\delta_{CP}} )
&s_{23}c_{13} \\
s_{23}s_{12}-c_{23}c_{12}s_{13} e^{i\delta_{CP}}
 & (-s_{23}c_{12}-c_{23}s_{12}s_{\nu13} 
e^{i\delta_{CP}}) 
& c_{23}c_{13}
\end{array} \right) \ 
$$
$$
\times {\rm diag}( 1, e^{i\beta}, e^{i\gamma}) ,
\eqno(3.4)
$$
where $c_{ij}=\cos \theta_{ij}$ and $s_{ij}=\sin \theta_{ij}$ with the mixing angles $\theta_{ij}$. 
Note that since
the Majorana neutrino fields  have no freedom of rephasing 
invariance, so that we can use only the rephasing freedom 
of the charged lepton fields to transform the form of $U$ to $U_\nu$.
Hereafter, we call the mixing matrix (3.4) as the 
Maki-Nakagawa-Sakata (MNS) mixing matrix \cite{MNS}.

The MNS mixing angles $\theta_{12}$, $\theta_{23}$ 
and $\theta_{13}$ are calculable from mixing matrix $U_\nu$ as,
$$
\sin^2 \theta_{12} = |U_{12}|^2/(1 - |U_{13}|^2), \ \ \ 
\sin^2 \theta_{23} =|U_{23}|^2/(1 - |U_{13}|^2), \ \ \ 
\sin^2 \theta_{13} = |U_{13}|^2. \\
\eqno(3.5)
$$
The $CP$-violating phase $\delta_{CP}$, the additional Majorana 
phase $\beta$ and $\gamma$  
in the representation Eq.(3.4) are also calculable 
and obtained as
$$
\delta_{CP}  = \mbox{arg}
          \left[
             \frac{U_{12}U_{22}^*}{U_{13}U_{23}^*} + 
             \frac{|U_{12}|^2}{1-|U_{13}|^2}
          \right]  ,
 \eqno(3.6)
 $$
 $$
\beta  = \mbox{arg} \left( \frac{U_{12}}{U_{11}}\right) , 
\ \ \ \ 
 \gamma   = \mbox{arg} \left( \frac{U_{13}}{U_{11}}e^{i\delta_{CP}}\right) .
 \eqno(3.7)
$$

The ratio of neutrino mass square differences $R_{\nu}$ 
is also calculable from the mass eigenvalues of $M_\nu$ as 
$$
R_{\nu} \equiv \frac{\Delta m_{21}^2}{\Delta m_{32}^2}
=\frac{m_{2}^2 -m_{1}^2}{m_{3}^2 -m_{2}^2} .
\eqno(3.8)
$$
The effective neutrino mass $\langle m \rangle$ 
of the neutrinoless double beta decay is  defined  by 
$$
\langle m \rangle=|m_1 U_{11}^2+m_2 U_{12}^2+m_3 U_{13}^2| .
\eqno(3.9)
$$

At present, we have the four observed values\cite{NuFIT}  as
$$
R_\nu \equiv \frac{m_2^2 -m_1^2}{m_3^2-m_2^2} 
= \frac{7.39^{+0.21}_{-0.20} \times 10^{-5} {\rm eV}^2}{
2.525^{+0.033}_{-0.031} \times 10^{-3} {\rm eV}^2} 
=(2.93 \pm{0.12} )\times 10^{-2} ,
\eqno(3.10)
$$
$$
\sin^2 \theta_{12} = 0.310^{+0.013}_{-0.012}, \ \ \ 
\sin^2 \theta_{13} = 0.02240^{+0.00065}_{-0.00066}, \ \ \ 
\sin^2 \theta_{23} = 0.582^{+0.015}_{-0.019}.
\eqno(3.11)
$$

Let us give our strategy of parameter fitting to the observables.
In our model we have three parameters, $|a_{eff}|$, $\alpha$, and $\xi$ in $M_\nu$.  
Therefore the MNS mixing angles $\theta_{12}$, $\theta_{23}$, 
and $\theta_{13}$  and the ratio of neutrino mass square 
differences $R_{\nu}$ and so on are functions of 
$|a_{eff}|$, $\alpha$, and $\xi$ in this model. 

The values of our three free parameters can be fixed by using  
the three experimental values (center values)  of $R_\nu$, 
$\sin^2 \theta_{12}$ and $\sin^2 \theta_{13}$ in (3.10) and (3.11) as follows:  
$$
\mbox{  \quad} a_{eff} =-4.069, \ \ \ \alpha = \mp 0.408^o, \ \ \ 
 \xi = 0.756  .
\eqno(3.12)
$$
There are two solutions $\alpha=-0.408$ and $\alpha=+0.408^o$. 
Hereafter, we call Case (A) for  $\alpha=-0.408^o$ and 
Case (B) for  $\alpha=+0.408^o$. 
The observed values versus parameter choices are given in
Table~1.   

\begin{table}[ht]
\caption{Observed values vs. our parameter choices} 
 
\label{table1}
\hspace*{-6mm}
\begin{center}
\begin{tabular}{|c|ccc|} \hline
   & $\sin^2 \theta_{12}$ &  $\sin^2 \theta_{13}$ & 
 $R_{\nu}\ [10^{-2}]$ \\ \hline
Obs & $0.310$   &  $0.02240$ &   $2.93 $ \\ 
    & $^{+0.013}_{-0.012}$ &  $^{+0.00065}_{-0.00066}$ & $ \pm 0.12 $\\ 
 Our choice & $0.314$ &  $0.02242$ &  $2.91$ \\
\hline 
\end{tabular}
\end{center}
\end{table}


For the input parameters in cases (A) and (B), we predict 
$$
\mbox{ \quad } 
\sin^2 \theta_{23} = 0.672, \ \ \ \delta_{CP} = \mp 119^o , 
\ \ \ \beta= \pm 10.8^o , \ \ \ \gamma = \pm 7.15^o .
\eqno(3.13)
$$
When we use  
$\Delta m_{32}^2=m_3^2-m_2^2=2.525 \times 10^{-3} {\rm eV}^2$,
we obtain the prediction of the neutrino mass $m_i$ and 
effective neutrino mass $\langle m \rangle$ for the both 
 (A) and (B) as follows: 
$$
m_1 = 0.0486 \mbox{\,\,eV}, \ \ \ m_2 = 0.0494 \mbox{\,\,eV}, 
\ \ \, m_3= 0.0698 \mbox{\,\,eV}, \ \ \ 
\langle m \rangle = 0.0328 \mbox{\,\,eV}.
\eqno(3.14)
$$
The predicted  values of  observables are listed in  Table 2.

In this three parameter model,  the three  
parameters are fixed so as to reproduce the observed values 
of  $R_\nu$, $\sin^2 \theta_{12}$ 
and $\sin^2 \theta_{13}$. 
The predictions of the $\delta_{CP}$, $\beta$, and $\gamma$ 
in (3.13) and of the neutrino masses $m_i$ 
and effective neutrino mass $\langle m \rangle$ in (3.14) 
will be checked in near future experiments.

We have constructed the neutrino mass matrix model 
with the number of free parameters as small as possible 
in order to have high predictability in our unified 
description approach for the quarks and leptons.  
It seems impossible to build a neutrino mass matrix model  
with furthermore few parameters, e.g. a two parameter model.

\begin{table}[ht]
\caption{Predicted values vs. observed values. 
} 
\label{table1}
\hspace*{-6mm}
\begin{center}
\begin{tabular}{|c|cccccccc|} \hline
   &  
$\sin^2 \theta_{23}$ & $\delta_{CP}$ & $\beta$ & $\gamma$ & $m_{1}\ [{\rm eV}]$ & $m_{2}\ [{\rm eV}]$ & 
$m_{3}\ [{\rm eV}]$ & $\langle m \rangle \ [{\rm eV}]$  \\ \hline
 Pred & $0.672$ & $\mp 119^\circ$& $\pm 10.8^\circ$ & $\pm 7.15^\circ$ &
 $0.0486$ & $0.0494$ & $0.0698$ & $0.0328$  \\
Obs  &  $0.582$  &  -  &  - &  -  &  -  &  -  &  -  &  $<\mathrm{O}(10^{-1})$\\
  & $^{+0.015}_{-0.019}$ & & & & & &  & \\ \hline 
\end{tabular}
\end{center}
\end{table}


\newpage

{\large\bf Acknowledgment}

The work was supported by JSPS KAKENHI Grant
number JP16K05325 (Y.K.).



\vspace{5mm}
%


\end{document}